\documentclass[11pt, a4paper, fleqn]{article}
\usepackage{amsmath}
\usepackage{amsfonts}
\usepackage{amsgen}
\usepackage{amssymb}
\usepackage{amsbsy}
\usepackage{graphicx}
\usepackage{appendix}
\usepackage{fancyhdr}
\usepackage{lastpage}
\usepackage{enumerate}
\usepackage{textcomp}

\newcommand{\aderv}[2]{\frac{\partial {#1}}{\partial {#2}}}

\newcommand{\equa}[1]{Eq.~(\ref{#1})}
\newcommand{\equas}[1]{Eqs.~(\ref{#1})}
\newcommand{\equass}[2]{Eqs.~(\ref{#1})-(\ref{#2})}
\newcommand{\equasa}[2]{Eqs.~(\ref{#1}){ }and{ }(\ref{#2})}
\newcommand{\eqn}[2]{\begin{equation}
#1
\label{#2}
\end{equation}
}
\newcommand{\spl}[2]{\begin{equation}
\begin{split}
#1
\end{split}
\label{#2}
\end{equation}
}

\title{\bf Local energy balance, specific heats and the Oberbeck-Boussinesq approximation}

\author{A.\ Barletta\\
\textit{\small $^{a}$Dipartimento di Ingegneria Energetica, Nucleare e del Controllo}\\
\textit{\small Ambientale (DIENCA), Universit\`{a} di Bologna, Via dei Colli 16, I--40136 Bologna, Italy}\\
\texttt{\small antonio.barletta@unibo.it}\\
}
\date{~}

\begin{document}
\linespread{1.4}

\maketitle

\begin{abstract}
\noindent A thermodynamic argument is proposed in order to discuss the most appropriate form of the local energy balance equation within the Oberbeck-Boussinesq approximation. The study is devoted to establish the correct thermodynamic property to be used in order to express the relationship between the change of internal energy and the temperature change. It is noted that, if the fluid is a perfect gas, this property must be identified with the specific heat at constant volume. If the fluid is a liquid, a definitely reliable approximation identifies this thermodynamic property with the specific heat at constant pressure. No explicit pressure work term must be present in the energy balance. The reasoning is extended to the case of fluid saturated porous media.

\end{abstract}

\vfill
\begin{center}
\noindent {\small \textbf{Keywords} --- Oberbeck-Boussinesq approximation; Energy balance; Specific heat; Porous media\\[-6pt] }
\vspace{100pt}
Paper submitted to the\\ \textsc{International Journal of Heat and Mass Transfer}
\end{center}
\vfill

\eject

\section{Introduction}
The Oberbeck-Boussinesq approximation, so named after the pioneering works by Oberbeck \cite{Ob} and Boussinesq \cite{Bo}, is the basis of most of the contemporary studies on natural or mixed convection flows. Very interesting historical surveys on the origins of this approximation are available in the recent papers by 
Zeytounian \cite{Zh} and Bois \cite{Bois}. 

Although the nature of this approximation is very clear and unambiguous with reference to the mass and momentum balance equations for the fluid, the formulation of the approximated energy balance equation is not so definite and univocal. The questions concerning the energy balance equations are the following:
	\newcounter{Lcount}
  \begin{list}{\Alph{Lcount})~~~}{\usecounter{Lcount}}
  \item Which is the specific heat involved in the energy balance?
  \item Is there a pressure work term in the energy balance, proportional to the convective derivative of the pressure field?
  \end{list}
The textbooks on fluid dynamics and heat transfer generally give clear answers to these questions. The problem, as it will be discussed in Section \ref{enbal}, is that the answers are different. 

In a recent technical note \cite{Bar}, the present author carried out a first analysis of the existing formulations of the local energy balance adopted in the Oberbeck-Boussinesq approximation of buoyant flows. 

The purpose of this short paper is to extend the analysis performed in Ref. \cite{Bar} in order to point out the manifold nature of the energy balance formulations, within the framework of the Oberbeck-Boussinesq approximation, available in the literature. Then, a thermodynamic argument is proposed in order to give answers to questions A) and B). In particular, it will be concluded that the answer to question A) depends on the fluid being a liquid or a gas. For a perfect gas the answer to question A) is definitely: ``the specific heat at constant volume $c_v$''. For a liquid the answer to question A) is less definite, but sufficiently reliable: ``the specific heat at constant pressure $c_p$''. The answer to question B) is: ``no pressure work term appears in the energy balance''. In a final section, the analysis of the energy balance is applied to the topic of buoyant flows in fluid saturated porous media.

\section{Mass and momentum balance}
As is well known, the Oberbeck-Boussinesq approximation implies that the local mass and momentum balance equations be written as
\eqn{\aderv{{u}_i}{{x}_i} = 0,}{1}
\eqn{\frac{\textrm{D} {u}_i}{\textrm{D} {t}} = - \frac{1}{{\rho}_0} \; \aderv{{p_e}}{{x}_i} - \left( {T} - {T}_0 \right) \beta\, g_i + \nu \, {\nabla}^2 {u}_i ,}{2}
where the summation over repeated indices is assumed. In \equa{2}, $\textrm{D}/\textrm{D} {t}$ is the substantial or convective derivative. In \equasa{1}{2}, $u_i$ is the velocity field, $x_i$ is the position vector, $t$ is the time, $T$ is the temperature, $g_i$ is the gravitational acceleration, $\nu$ is the kinematic viscosity, $\beta$ is the isobaric coefficient of thermal expansion, ${\rho}_0$ and ${T}_0$ are the reference density and the reference temperature respectively, while $\textrm{D}/\textrm{D} {t}$ is the substantial or convective derivative. In \equasa{1}{2}, the properties $\nu$ and $\beta$ are also referred to the temperature $T_0$. In \equasa{1}{2}, the properties $\nu$ and $\beta$ are referred to the temperature $T_0$. The implicit assumptions behind \equasa{1}{2} are that: ${p_e} = p - \rho_0\, g_i\, x_i$ is the difference between the pressure $p$ and the hydrostatic pressure, and that one considers the density as coincident with the reference value ${\rho}_0$ except for the gravitational body force term. For that term, the density $\rho$ is assumed to be a function of the temperature only, thus considering the dependence on the pressure as negligible. The linear equation of state
\eqn{{\rho} \left( {T} \right) = {\rho}_0 \left[ 1 - \beta \left( {T} - {T}_0 \right) \right] 
}{3}
is implicitly invoked in \equa{2}.
In \equa{3}, the dependence on ${T}$ is assumed to be sufficiently weak to be approximated linearly in the surroundings of the reference value ${T}_0$. 

\section{Energy balance}
\label{enbal}
For the energy balance, the formulation of the Oberbeck-Boussinesq approximation is not so definite in the literature. In fact, one may have Chandrasekhar's \cite{Ch} and White's \cite{Wh} formulation
\eqn{{\rho}_0 \; c_v\; \frac{\textrm{D} {T}}{\textrm{D} {t}} = k\, {\nabla}^2 {T}  + 2\, \mu\, {\mathfrak{D}}_{i j}\, {\mathfrak{D}}_{i j}  ,
}{4}
where $k$ is the thermal conductivity, $\mu$ is the dynamic viscosity and ${\mathfrak{D}}_{i j}$ is the strain tensor
\eqn{{\mathfrak{D}}_{i j} = \frac{1}{2} \left( \aderv{{u}_j}{{x}_i} + \aderv{{u}_i}{{x}_j} \right) .}{5}
The source term in \equa{5}, $2\, \mu\, {\mathfrak{D}}_{i j}\, {\mathfrak{D}}_{i j}$, is the thermal power generated by the viscous dissipation.

One may have the enthalpy formulation \cite{MK,Tu}
\eqn{{\rho}_0 \; c_p\; \frac{\textrm{D} {T}}{\textrm{D} {t}} = k\, \nabla^2 {T}  + 2\, \mu\, {\mathfrak{D}}_{i j}\, {\mathfrak{D}}_{i j} + \beta \, {T} \; \frac{\textrm{D} {p}}{\textrm{D} {t}}  ,
}{6}
where the last term on the right hand side is an additional source term: the pressure work acting on the fluid element.

Finally, one may have Landau-Lifshitz's \cite{LL}, Bejan's \cite{Be} and Kundu-Cohen's \cite{KK} formulation
\eqn{{\rho}_0 \; c_p\; \frac{\textrm{D} {T}}{\textrm{D} {t}} = k\, \nabla^2 {T}  + 2\, \mu\, {\mathfrak{D}}_{i j}\, {\mathfrak{D}}_{i j} .
}{7}
In order to decide on the most convenient expression of the energy balance, let us write the general non-approximated form of this balance \cite{Ch, AL}, \emph{i.e.} the local version of the First Law of thermodynamics
\eqn{{\rho} \; \frac{D {e}}{D {t}} = -\, \aderv{{q}_i}{{x}_i} + {\sigma}_{i j}\, {\mathfrak{D}}_{i j}  ,
}{8}
where ${e}$ is the internal energy per unit mass, ${\sigma}_{i j}$ is the fluid stress tensor and ${q}_i = -\, k \, \partial {T} / \partial {x}_i$ is the heat flux density. The meaning of the terms on the right hand side of \equa{8} is straightforward: the first term is the incoming heat flux contribution to the energy change, while the second term is the mechanical work input due to the forces acting on the boundary of the fluid element. The latter term depends only on mechanical quantities,\emph{ i.e.} on the velocity and pressure fields within the fluid domain. On the other hand, the evaluation of the energy change on the left hand side of \equa{8} implies the use of thermodynamics. 

Thermodynamics ensures that ${e} = {e}\left( {T}, {\rho} \right)$ for every single-phase or two-phase stable equilibrium states. In the special case of a perfect gas, it is well known that ${e} = {e}\left( {T} \right)$ \cite{He}, so that 
\eqn{d {e} = c_v \, d {T} .
}{9}
In the case of either a liquid or a real gas, one must rely on the Oberbeck-Boussinesq approximation by assuming that an approximate equation of state ${\rho} = {\rho} \left( {T} \right)$ can be applied. This implies that the pressure of the fluid does not change appreciably. Since ${\rho} = {\rho} \left( {T} \right)$ and since the pair $\left( {T}, {\rho} \right)$ yields a unique stable equilibrium state, then one concludes that all the thermodynamic properties may be considered as functions of ${T}$. This conclusion holds for the internal energy per unit mass, so that a relationship
\eqn{d {e} = c \, d {T} 
}{10}
can be established. The thermodynamic coefficient $c$, in general, does not coincide either with $c_v$ or with $c_p$. In fact, $c$ is the total derivative of the function ${e} = {e}\left( {T}, {\rho} \left( {T} \right) \right)$ with respect to ${T}$, and not the partial derivative of ${e} = {e}\left( {T}, {\rho} \right)$ with ${\rho}$ kept constant. As is well known, the latter is the correct thermodynamic definition of $c_v$ \cite{He}. The equation of state ${\rho} = {\rho} \left( {T} \right)$ is one regarding a set of stable equilibrium states of the fluid with approximately the same pressure. Then, one has
\eqn{c = \left( \aderv{e}{T} \right)_p . 
}{11}
\equa{11} is not the definition of the specific heat at constant pressure $c_p$. As is well known \cite{He}, the latter is defined as
\eqn{c_p = \left( \aderv{h}{T} \right)_p , 
}{12}
where $h = e + p/\rho$ is the enthalpy per unit mass. Then, one can easily write the following relationship:
\eqn{c = c_p - \frac{p\, \beta}{\rho} ,
}{13}
where the definition of the coefficient of isobaric expansion
\eqn{\beta = - \, \frac{1}{\rho} \left( \aderv{\rho}{T} \right)_p
}{14}
has been used.
Then, $c$ is smaller than $c_p$ and differs from $c_v$, except for the limiting case of a perfect gas. Indeed, in the latter case, one can easily show that the equation of state of the perfect gas and \equa{13} ensure that $c = c_v$, so that \equasa{9}{10} are perfectly consistent. 

With reference to a Newtonian fluid and on account of \equa{1}, one may express ${\sigma}_{i j}\, {\mathfrak{D}}_{i j} = 2\, \mu\, {\mathfrak{D}}_{i j}\, {\mathfrak{D}}_{i j}$. 
Then, by employing \equa{10}, by replacing ${\rho}$ with ${\rho}_0$ and by assuming $k = constant$, \equa{8} can be approximated as
\eqn{{\rho}_0 \, c\; \frac{\textrm{D} {T}}{\textrm{D} {t}} = k\, {\nabla}^2 {T}  + 2\, \mu\, {\mathfrak{D}}_{i j}\, {\mathfrak{D}}_{i j}  .
}{15}
In \equa{15}, the value of the thermodynamic coefficient $c$, given by \equa{13}, is referred to the temperature $T_0$ like any other fluid property involved in the Oberbeck-Boussinesq system of governing equations. \equa{15} coincides with \equa{4} only for the limiting case of a perfect gas and differs from \equa{7} as $c < c_p$. 
\begin{figure}[t]
	\centering
		\includegraphics[height=0.4\textwidth]{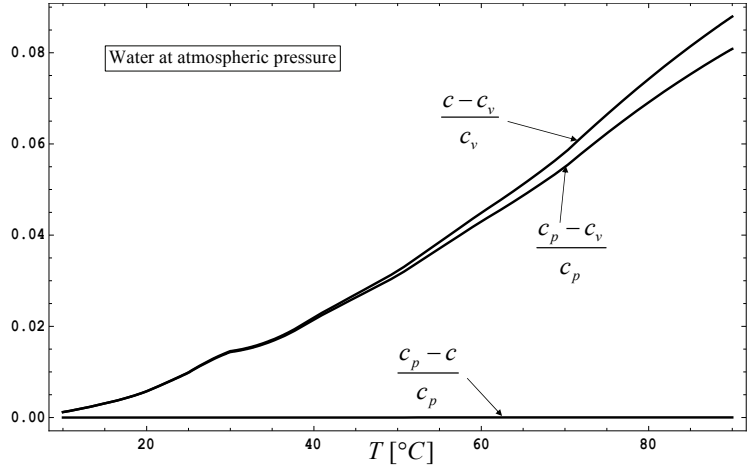}
	\caption{Comparison of $c$ with $c_p$ and $c_v$ for water at atmospheric pressure}
	\vspace{10pt}
	\label{fig1}
\end{figure}
\begin{figure}[t]
	\centering
		\includegraphics[height=0.4\textwidth]{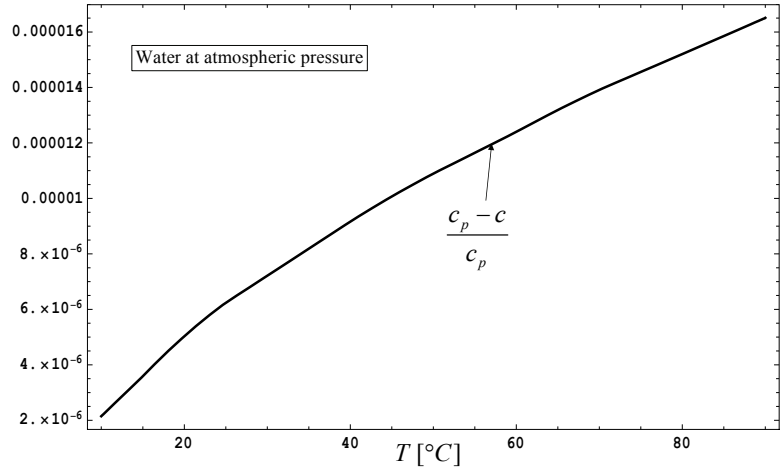}~~~~
	\caption{Comparison of $c$ with $c_p$ for water at atmospheric pressure}
	\vspace{10pt}
	\label{fig2}
\end{figure}

One can question the extent to which $c_p$ and $c_v$ differ from $c$ in the case of liquids. For water at atmospheric pressure, a precise evaluation of the discrepancies $(c_p - c) / c_p$, $(c - c_v) / c_v$ and $(c_p - c_v) / c_p$ can be done by means of the data reported in Appendix C of Bejan's textbook \cite{Be}. In fact, on account of \equa{13}, one has
\eqn{\frac{c_p - c}{c_p} = \frac{p\, \beta}{\rho \, c_p} , \qquad \frac{c - c_v}{c_v} = \frac{\rho \left( c_p - c_v \right) - p\, \beta}{\rho \, c_v } = \frac{\rho \left( c_p - c_v \right) - p\, \beta}{\rho \, c_p - \rho \left( c_p - c_v \right) } .
}{16}
Figure \ref{fig1} evidently suggests that the discrepancy $(c - c_v) / c_v$ is rather close to $(c_p - c_v) / c_p$ and that $(c_p - c) / c_p$ is much smaller. Figure \ref{fig2} shows that $(c_p - c) / c_p$ is, on average, of order $10^{-5}$. As a consequence, for water at atmospheric pressure, the approximation
\eqn{c \cong c_p 
}{17}
is a definitely reliable one. In general, it is not easy to find data for the specific heat at constant volume of a liquid. Usually, thermodynamic tables report the values of $c_p$, while $c_v$ is evaluated from the Mayer relationship \cite{He}
\eqn{c_p - c_v = \frac{\beta^2\, T}{\rho\, \kappa_T},
}{18}
where $\kappa_T$ is the coefficient of isothermal compressibility,
\eqn{\kappa_T = \frac{1}{\rho} \left( \aderv{\rho}{p} \right)_T .
}{19}
On account of \equasa{16}{18} and of the data reported in Refs.  \cite{He, CRC}, one may evaluate the discrepancies $(c_p - c) / c_p$, $(c - c_v) / c_v$ and $(c_p - c_v) / c_p$ for some organic liquids at $25^\circ \textrm{C}$ and atmospheric pressure. These data are reported in Table \ref{tab:tab1}. This table suggests again that \equa{17} is definitely reliable with an error smaller than $0.009\%$, while the approximation $c \cong c_v$ would lead to an error higher than $35 \%$. 
\begin{table}[t]
	\caption{Specific heat ratios for some organic liquids at $25^\circ \textrm{C}$ and atmospheric pressure}
	\begin{center}
\begin{tabular}{l |  c  c  c }
 & $(c_p - c) / c_p$ & $(c - c_v) / c_v$ & $(c_p - c_v) / c_p$  \\
\hline
Acetone ($\mathrm{C_3 H_6 O}$)		& 0.0087\% & 40\% & 29\%  \\
Benzene ($\mathrm{C_6 H_6}$)			& 0.0075\% & 36\% & 26\%  \\
Ethanol ($\mathrm{C_2 H_6 O}$)		& 0.0074\% & 37\% & 27\%  \\
Methanol ($\mathrm{C H_4 O}$)			& 0.0076\% & 37\% & 27\%  \\
\hline
\end{tabular}
\end{center}
\vspace{10pt}	%
	\label{tab:tab1}
\end{table}

\section{Possible pitfalls}
In the preceding section, a thermodynamic strategy has been established to determine the most appropriate formulation of the local energy balance equation within the Oberbeck-Boussinesq approximation. The basis of this approach is twofold. 
\begin{itemize}
	\item The convective derivative of the internal energy per unit mass, $\textrm{D} e / \textrm{D} {t}$, is evaluated by considering the thermodynamic process undergone by the fluid element. This process can be reliably modeled as an isobaric process.
	\item The mechanical work input ${\sigma}_{i j}\, {\mathfrak{D}}_{i j}$ is evaluated according to the stress-strain relationship for a Newtonian fluid, as well as to the constraint $\partial u_i/\partial x_i = 0$ satisfied by the velocity field.  
\end{itemize}
In the light of these arguments, three possible pitfalls that can be encountered in the determination of the local energy balance equation are described in the following.

\subsection{The isochoric process}
\label{isocho}
One could say that the convective derivative $\textrm{D} e / \textrm{D} {t}$ can be evaluated by assuming that the fluid element is undergoing an isochoric thermodynamic process \cite{Ch}. Therefore the validity of \equa{9} is extended, not only to the perfect gases, but to every fluid system. Then one would be led to \equa{4} instead of \equa{15}. This reasoning is erroneous as the nature of the thermodynamic process undergone by the fluid element is implicitly defined by the equation of state assumed, \emph{i.e.} \equa{3}. This equation of state is based on the hypothesis that the density may change only as a consequence of temperature variations. In other words, one is assuming that the set of stable equilibrium states available to the fluid element are, with a very good approximation, at constant pressure.

\subsection{The expansion-contraction work}
A possible misleading argument in the deduction of \equa{15} is connected to the mechanical work term ${\sigma}_{i j}\, {\mathfrak{D}}_{i j}$. This term, as it is clearly explained in Chandrasekhar \cite{Ch}, must be simplified according to the constraint satisfied by the velocity field, \emph{i.e.} \equa{1}. On the other hand, in some textbooks (see, for instance, Kundu and Cohen \cite{KK}), a part of the mechanical work term ${\sigma}_{i j}\, {\mathfrak{D}}_{i j}$, namely the expansion-contraction work contribution, $- p\, \partial u_i/\partial x_i$, is rewritten by forgetting \equa{1} and by using the exact local mass balance instead,
\eqn{- p\; \aderv{{u}_i}{{x}_i} = \frac{p}{\rho}\; \frac{\textrm{D} \rho}{\textrm{D} {t}}.
}{20c}
If one brings this term to left hand side of the local energy balance \equa{8}, then one has to evaluate
\eqn{\rho\; \frac{\textrm{D} e}{\textrm{D} {t}} - \frac{p}{\rho}\; \frac{\textrm{D} \rho}{\textrm{D} {t}}
}{21c} 
instead of $\rho\, \textrm{D} e / \textrm{D} {t}$. Therefore, by following the thermodynamic argument described in Section \ref{enbal}, one would have
\eqn{\rho\, d e - \frac{p}{\rho} \; d \rho = \rho\, c \, d T + p \, \beta \, d T = \rho \left( c_p - \frac{p\, \beta}{\rho} \right) d T + p \, \beta \, d T = \rho\, c_p \, d T,
}{22c}
where \equas{10}, (\ref{13}) and (\ref{14}) have been used. On account of \equass{20c}{22c}, one would be led to the formulation of the local energy balance expressed by \equa{7}, instead of \equa{15}. However, this is a tricky procedure, since the velocity field evaluated through the Oberbeck-Boussinesq model is constrained to be solenoidal by \equa{1}. Then, even if it is based on an exact mass balance, \equa{20c} cannot be coherently invoked within the Oberbeck-Boussinesq approximation. 

\subsection{The pressure work}
One could evaluate the differential of the internal energy per unit mass, $d e$, by using the definition of enthalpy per unit mass, $h = e + p/\rho$. Then, one has
\eqn{\rho \, d e = \rho \, d h -  d p + \frac{p}{\rho}\;d \rho .
}{20b}

If one assumes that the thermodynamic process undergone by the fluid element is isobaric, then one has
\eqn{d p = 0 , \quad d h = c_p \, d T, \quad d \rho = - \rho\, \beta\, d T.
}{21b}
Then, \equa{20b} yields
\eqn{\rho \, d e = \rho \, c_p \, d T - p\, \beta\, d T = \rho \left( c_p - \frac{p\, \beta}{\rho} \right) d T = \rho \, c \, d T,
}{22b}  
where \equa{13} has been used. \equa{22b} leads directly to \equa{15}.

If one assumes that the thermodynamic process undergone by the fluid element is isochoric, then one has
\eqn{d \rho = 0 , \quad d h = c_p \, d T + \left( \aderv{h}{p} \right)_T d p .
}{23b}
The thermodynamic identity 
\eqn{\rho \left( \aderv{h}{p} \right)_T = 1 - \beta\, T
}{24b}
can be easily proved on the basis of the elementary thermodynamic differential relationships \cite{He}. The complete proof can be found, for instance, in Ref. \cite{AL}. Therefore, by substituting \equasa{23b}{24b} into \equa{20b}, one obtains 
\eqn{\rho \, d e = \rho \, c_p \, d T + \left( 1 - \beta\, T \right) d p -  d p = \rho \, c_p \, d T - \beta\, T \, d p .
}{25b}
From \equa{25b}, one justifies the relationship
\eqn{{\rho} \; \frac{\textrm{D} {e}}{\textrm{D} {t}} = {\rho} \, c_p\; \frac{\textrm{D} {T}}{\textrm{D} {t}} - \beta\, T \; \frac{\textrm{D} {p}}{\textrm{D} {t}} .
}{26b}
Obviously, \equa{26b} leads to \equa{6}.

While the procedure based on \equass{20b}{22b} and leading to \equa{15} is correct, the procedure based on \equas{20b}, (\ref{23b})-(\ref{26b}) and leading to \equa{6} is completely unjustified for the following two reasons.
\begin{itemize}
	\item The process undergone by the fluid element cannot be considered as isochoric as already pointed out in Subsection \ref{isocho}.
	\item If one has to suppose that the process is isochoric, the straightforward reasoning is to assume the general validity of \equa{9}, so that one is led to \equa{4}, as described in Subsection \ref{isocho}.
\end{itemize}
In other words, the misleading thermodynamic analysis behind \equa{6} adds an unjustified assumption (an isochoric process) to a tricky procedure that replaces the simple conclusion
\eqn{{\rho} \; \frac{\textrm{D} {e}}{\textrm{D} {t}} = {\rho} \, c_v\; \frac{\textrm{D} {T}}{\textrm{D} {t}}  
}{27b}
with the more complicated one expressed by \equa{26b}. This circumstance induces some reflections on how the erroneous assumption of isochoric process can lead to ambiguous results, depending on the procedure followed. On the contrary, the assumption of isobaric process leads exactly to the same result, \emph{i.e.} \equa{15}, either if the reasoning stems from the evaluation of $d e$ or, as in the present subsection, if one works out the differential of the enthalpy per unit mass, $d h$.

\section{On porous media}
The remarks on the appropriate form of the local energy balance in the framework of the Oberbeck-Boussinesq approximation can be easily reformulated with reference to the theory of fluid saturated porous media. In fact, following Nield and Bejan \cite{NB}, the volume-averaged energy equations for the solid and fluid phases
can be written as
\eqn{ \left( 1 - \varphi \right) (\rho\, c_v)_s \aderv{T_s}{t} = \left( 1 - \varphi \right) k_s \nabla^2 T_s + H \left( T_f - T_s \right) ,
}{20}
\eqn{ \varphi \, (\rho\, c)_f \aderv{T_f}{t} + (\rho\, c)_f \, V_i \aderv{T_f}{x_i} = \varphi \, k_f  \nabla^2 T_f + \mu \, \Phi + H \left( T_s - T_f \right),
}{21}
where $V_i$ is the seepage velocity and the subscripts $s$ and $f$ denote the solid and the fluid phase, respectively. In \equa{21}, $\varphi$ is the porosity, while $\mu\, \Phi$ is the viscous dissipation term obtained through a volume average of the term $2\, \mu\, {\mathfrak{D}}_{i j}\, {\mathfrak{D}}_{i j}$ appearing in \equa{15}. As is well known \cite{BR, N0}, the specific form of the term $\mu\, \Phi$ depends on the momentum balance model adopted for the fluid saturated porous medium. The inter-phase heat transfer coefficient $H$ in \equasa{20}{21} describes the thermal energy flow between the solid and the fluid phase. Since reference is made to the Oberbeck-Boussinesq approximation, all the solid and fluid properties in \equasa{20}{21} are evaluated at the reference temperature $T_0$. With respect to Nield and Bejan \cite{NB}, \equa{21} has been adapted on the basis of \equa{15} in order to include the thermodynamic coefficient $c$ defined through \equa{13}. In cases of local thermal equilibrium between the solid phase and the fluid phase, $T_s = T_f = T$, one can add \equasa{20}{21}, so that one has 
\eqn{ (\rho\, c)_m \aderv{T}{t} + (\rho\, c)_f \, V_i \aderv{T}{x_i} = k_m  \nabla^2 T + \mu \, \Phi ,
}{22}
where
\spl{ 
& (\rho\, c)_m =  \left( 1 - \varphi \right) (\rho\, c_v)_s + \varphi\, (\rho\, c)_f  , \\
& k_m =  \left( 1 - \varphi \right) k_s + \varphi\, k_f  . 
}{23}

\section{Conclusions}

The nature of the Oberbeck-Boussinesq approximation has been revisited in order to clarify some thermodynamic aspects connected to the formulation of the energy balance. The present analysis, motivated by the manifold formulation of this balance in the existing literature, leads to the following conclusions:
\begin{enumerate}
\item the energy balance for a fluid, within the Oberbeck-Boussinesq approximation, is given by the equation
\[
{\rho}_0 \, c\; \frac{\textrm{D} {T}}{\textrm{D} {t}} = k\, {\nabla}^2 {T}  + 2\, \mu\, {\mathfrak{D}}_{i 		j}\, {\mathfrak{D}}_{i j}  ,
\]
where $c = c_p - p\, \beta / \rho$;
\item the thermodynamic coefficient $c$ coincides with the specific heat at constant volume, $c_v$, for a 		perfect gas; 
\item the thermodynamic coefficient $c$ is definitely well approximated by the specific heat at constant 			pressure, $c_p$, for a liquid;
\item no pressure work term of the type $\beta \, {T} \, \textrm{D} {p}\,/\textrm{D} {t}$ must be introduced on the right hand side of the energy balance equation.
\end{enumerate}
Finally, the most appropriate form of the energy balance for buoyant flows in a fluid saturated porous medium has been discussed.

\section*{Acknowledgment}
I am deeply indebted with Prof. Don Nield for his encouragement throughout the course of this work, as well as for his very helpful and insightful comments.

\vfill
\eject

\end{document}